\documentclass[conference]{IEEEtran}

\AtBeginDocument{%
 \providecommand\BibTeX{{%
   \normalfont B\kern-0.5em{\scshape i\kern-0.25em b}\kern-0.8em\TeX}}}

\usepackage{colortbl}
\usepackage{xspace}
\newcommand{\ie}{\textit{i.e., \xspace}}
\newcommand{\eg}{\textit{e.g., \xspace}}
\newcommand{\etal}{\textit{et al. \xspace}}

\usepackage{array}
\newcolumntype{L}{>{\arraybackslash}m{16cm}}
\newcolumntype{C}[1]{>{\centering\let\newline\\arraybackslash\hspace{0pt}}m{#1}}
\newcolumntype{R}[1]{>{\raggedleft\let\newline\\arraybackslash\hspace{0pt}}m{#1}}
\usepackage[numbered]{bookmark}
\newcommand{\ballnumber}[1]{\tikz[baseline=(myanchor.base)] \node[circle,fill=.,inner sep=1pt] (myanchor) {\color{-.}\bfseries\footnotesize #1};}

\usepackage[many]{tcolorbox}
\usepackage{graphicx}
\usepackage[switch]{lineno}
\usepackage{fontawesome}
\usepackage{lineno}
\usepackage{xcolor}
\usepackage{stfloats}
\usepackage[justification=centering]{caption}
\usepackage{dirtytalk}
\usepackage{amsmath}
\usepackage{pgfplots, pgfplotstable}
\usetikzlibrary{patterns}
\usepackage{supertabular} 
\usepackage[flushleft]{threeparttable}
\usepackage{supertabular}

\usepackage[english]{babel}
\usepackage{textcomp}
\usepackage{mathtools}
\usepackage{ltxtable}
\usepackage{algorithm}
\usepackage{algpseudocode}
\usepackage{textcomp}

\usepackage{pgf-pie}
\definecolor{wedge1}{RGB}{ 190  30  46}
\definecolor{wedge2}{RGB}{ 240  65  54}
\definecolor{wedge3}{RGB}{ 241  90  43}
\definecolor{wedge4}{RGB}{ 247 148  30}
\definecolor{wedge5}{RGB}{  43  56 144}
\definecolor{wedge6}{RGB}{  28 117 188}
\definecolor{wedge7}{RGB}{  40 170 225}
\definecolor{wedge8}{RGB}{ 119 179 225}
\definecolor{wedge9}{RGB}{ 181 212 239}
\definecolor{wedge10}{RGB}{  0 104  56}
\definecolor{wedge11}{RGB}{  0 148  69}
\definecolor{wedge12}{RGB}{ 57 181  74}
\definecolor{wedge13}{RGB}{141 199  63}
\definecolor{wedge14}{RGB}{215 244  34}
\definecolor{wedge15}{RGB}{249 237  50}
\definecolor{wedge16}{RGB}{248 241 148}
\definecolor{wedge17}{RGB}{242 245 205}
\definecolor{wedge18}{RGB}{123  82  49}
\definecolor{wedge19}{RGB}{104  73 158}
\definecolor{wedge20}{RGB}{102  45 145}
\definecolor{wedge21}{RGB}{148 149 151}
\definecolor{wedge22}{RGB}{ 204 50 153}
\definecolor{wedge23}{RGB}{ 79 47 79}
\definecolor{wedge24}{RGB}{ 173 234 234}
\definecolor{wedge25}{RGB}{ 216 191 216}
\definecolor{wedge26}{RGB}{  43  56 144}
\definecolor{wedge27}{RGB}{  40 170 225}
\definecolor{wedge28}{RGB}{ 119 179 225}
\definecolor{wedge29}{RGB}{ 181 212 239}
\definecolor{wedge30}{RGB}{  0 104  56}
\definecolor{wedge31}{RGB}{  0 148  69}
\definecolor{wedge32}{RGB}{ 57 181  74}

\usetikzlibrary{chains}

\pgfplotstableread[col sep=comma,]{data.csv}\datatable

\pgfplotstablecreatecol[
    create col/expr={( \thisrow{B} ) / ( \thisrow{B} + \thisrow{C} + \thisrow{D} + \thisrow{E} + \thisrow{F} )}] {B_fraction} \datatable
\pgfplotstablecreatecol[
    create col/expr={( \thisrow{C} ) / ( \thisrow{B} + \thisrow{C} + \thisrow{D} + \thisrow{E} + \thisrow{F} )}] {C_fraction} \datatable
\pgfplotstablecreatecol[
    create col/expr={( \thisrow{D} ) / ( \thisrow{B} + \thisrow{C} + \thisrow{D} + \thisrow{E} + \thisrow{F} )}] {D_fraction} \datatable
\pgfplotstablecreatecol[
    create col/expr={( \thisrow{E} ) / ( \thisrow{B} + \thisrow{C} + \thisrow{D} + \thisrow{E} + \thisrow{F} )}] {E_fraction} \datatable
\pgfplotstablecreatecol[
    create col/expr={( \thisrow{F} ) / ( \thisrow{B} + \thisrow{C} + \thisrow{D} + \thisrow{E} + \thisrow{F} )}] {F_fraction} \datatable

\pgfmathsetmacro\startAngle{90-3.6/2}
\pgfmathsetmacro\radius{+5}
\pgfmathsetmacro\maxLeg{+12}
\pgfmathsetmacro\legBound{+60}
\pgfmathsetmacro\legSpacing{2*\legBound/(\maxLeg-1)}
  \pgfplotstableread[row sep=\\,col sep=&]{
        interval & carT & sd\\
     }\mydata

\usepackage{verbatim}
\pgfplotsset{compat=1.15}
\usetikzlibrary{patterns,}

\usepackage{graphicx}
\usepackage{tabularx,booktabs}
\usepackage{array}
\usepackage{bbding}
\usepackage{pifont}
\usepackage{wasysym}
\usepackage{caption}
\usepackage{dirtytalk}
\usepackage{subcaption}
\usepackage{tabularx}
\usepackage{sfmath}
\usepackage{array, booktabs, makecell}
\usepackage{color}
\usepackage{csquotes}
\usepackage{hyperref}
\usepackage{comment}
\usepackage{tikz}
\usepackage{balance}
\usepackage{listings}
\usepackage{booktabs} 
\usepackage{soul} 
\usetikzlibrary{fit} 
\usetikzlibrary{positioning}
\usetikzlibrary{arrows}
\usetikzlibrary{shapes.multipart}
\usepackage{adjustbox}
\usepackage{float}
\usepackage{rotating}
\usepackage{nth}
\DeclareCaptionType{TextBox}
\usepackage{pstricks}
\usepackage{placeins}
\usepackage{afterpage}
\usepackage{multirow} 
\usepackage{textcomp}
\usepackage{multicol}
\usepackage{varwidth} 

\definecolor{findOptimalPartition}{HTML}{D7191C}
\definecolor{storeClusterComponent}{HTML}{FDAE61}
\definecolor{dbscan}{HTML}{ABDDA4}
\definecolor{constructCluster}{HTML}{2B83BA}

\newcommand{\todo}[1]{\textcolor{red}{{\it [Added]}}}
\definecolor{main}{HTML}{5989cf}    
\definecolor{sub}{HTML}{cde4ff}     
\tcbset{
    sharp corners,
    colback = white,
    before skip = 0.2cm,    
    after skip = 0.5cm      
}   
\newtcolorbox{boxE}{
    enhanced, 
    boxrule = 0pt, 
    borderline = {0.75pt}{0pt}{main}, 
    borderline = {0.75pt}{2pt}{sub} 
}
\newtcolorbox{boxK}{
    sharpish corners, 
    boxrule = 0pt,
    toprule = 4.5pt, 
    enhanced,
    fuzzy shadow = {0pt}{-2pt}{-0.5pt}{0.5pt}{black!35} 
}
\usepackage{listings}

\definecolor{javared}{rgb}{0.6,0,0} 
\definecolor{javagreen}{rgb}{0.25,0.5,0.35} 
\definecolor{javapurple}{rgb}{0.5,0,0.35} 
\definecolor{javadocblue}{rgb}{0.25,0.35,0.75} 
 
\definecolor{light-red}{rgb}{1,0.92,0.91}
\definecolor{light-green}{rgb}{0.9,1,0.93}

\author{
\IEEEauthorblockN{Owen Truong\IEEEauthorrefmark{1},
Terrence Zhang\IEEEauthorrefmark{1},
Arnav  Marchareddy\IEEEauthorrefmark{1},
Ryan Lee\IEEEauthorrefmark{1},
Jeffery Busold\IEEEauthorrefmark{1}, \\
Michael Socas\IEEEauthorrefmark{1},
Eman Abdullah AlOmar\IEEEauthorrefmark{1}
}
\IEEEauthorblockA{\IEEEauthorrefmark{1}Stevens Institute of Technology, Hoboken, New Jersey, USA\\
\{otruong,tzhang68,amarchar,rlee16,jbusold,msocas,ealomar\}@stevens.edu\\
}} 

\newcommand{\toolname}{\textsc{LeakageDetector}\xspace}

\begin{document}


\title{LeakageDetector 2.0: Analyzing Data Leakage in Jupyter-Driven  Machine Learning Pipelines}


\maketitle
\begin{abstract}
In software development environments, code quality is crucial. This study aims to assist Machine Learning (ML) engineers in enhancing their code by identifying and correcting Data Leakage issues within their models. Data Leakage occurs when information from the test dataset is inadvertently included in the training data when preparing a data science model, resulting in misleading performance evaluations.  ML developers must carefully separate their data into training, evaluation, and test sets to avoid introducing Data Leakage into their code. In this paper, we develop a new Visual Studio Code (VS Code) extension, called \toolname, that detects Data Leakage — mainly Overlap, Preprocessing and Multi-test leakage — from Jupyter Notebook files. Beyond detection, we included two correction mechanisms: a conventional approach, known as a quick fix, which manually fixes the leakage, and an LLM-driven approach that guides ML developers toward best practices for building ML pipelines.  The plugin and its source code are publicly available on GitHub at \url{https://github.com/SE4AIResearch/DataLeakage\_JupyterNotebook\_Fall2024}. The demonstration video can
be found on YouTube: \url{https://youtu.be/7YiYVBiID_8}. The website can be found at \url{https://leakage-detector.vercel.app/}.
\end{abstract}

\begin{IEEEkeywords}
data leakage, machine learning, quality
\end{IEEEkeywords}



\section{Introduction}
\label{Section:Introduction} 

The ubiquity and invisibility of Data Leakage and the convenience of an automated solution are why we chose to extend \toolname \cite{alomar2025leakagedetector} to support Jupyter Notebooks. Data Leakage occurs when information from the test dataset is inadvertently included in the training data when preparing a data science model. It can lead to inaccurate predictions, distrust, and, at worst, legal issues. Thus,  having a tool dedicated solely to detecting it will be a major boon to data security. Data scientists, Machine Learning (ML) engineers, and statisticians can integrate the tool into their regular environments without issue. 

 Yang \etal \cite{yang2022data} have proposed a static analysis
approach to detect Data Leakage types (\ie Overlap, Preprocessing, and Multi-test) in ML code. Despite attempts to create an automated system for detecting Data Leakage, the tool is aimed at facilitating large-scale empirical investigations. Its adoption by professionals could be improved by addressing factors such as usability, user involvement, and correction of leakage. Therefore, by automatically detecting leakage and providing solutions, as well as showing the exact type of leakage and the variable that caused it, developers will gain a detailed understanding of what generally causes this type of leakage. The quick fixes offered by the tool double as teaching good coding practices, so if a developer uses the tool consistently for a long time, they could adopt the suggested coding practices habitually. 

\begin{figure*}[htbp]
  \centering
  \includegraphics[width=1.0\textwidth]{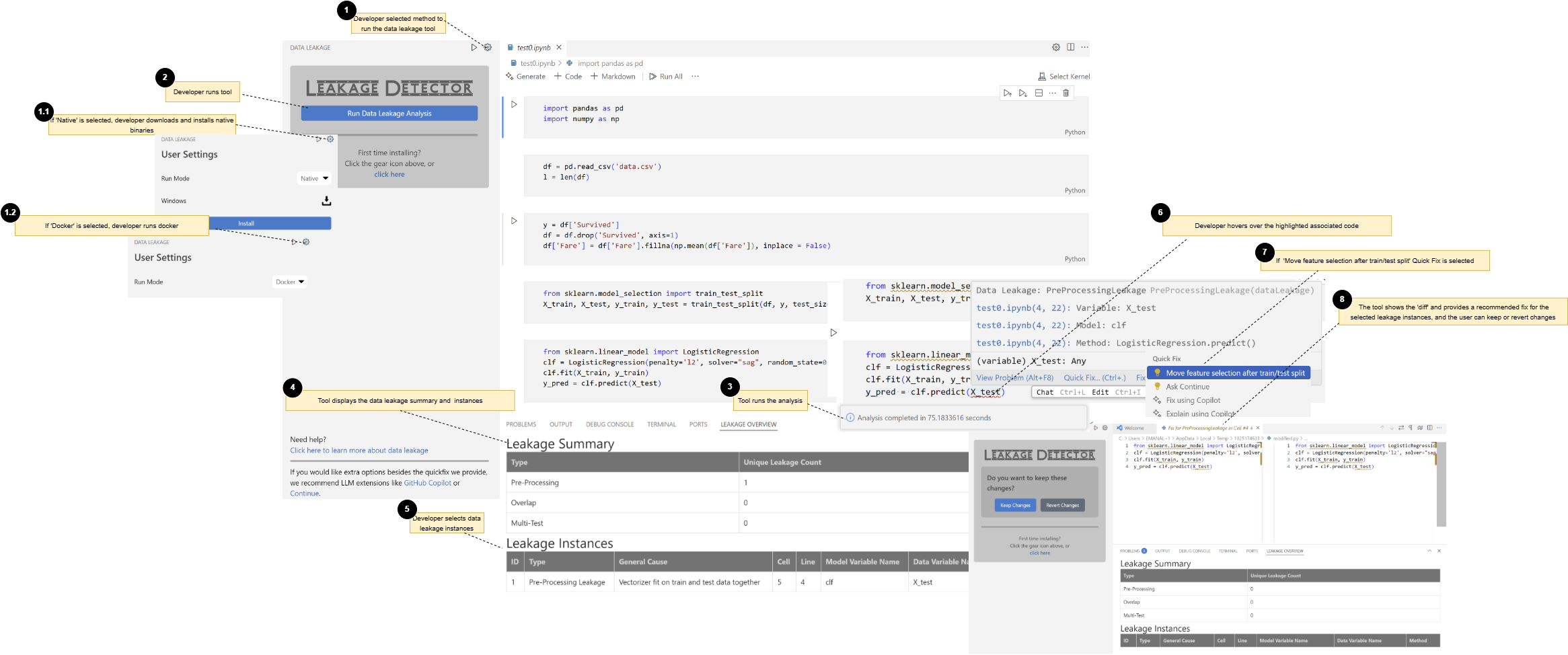}
  \caption{\toolname in action, showing the identified and fixed Data Leakage instances.}
  \label{fig:example}
  \vspace{-0.3cm}
 \end{figure*}
 
The initial release of the \toolname was developed to detect Data Leakage in Python files using PyCharm IDE as the primary platform, due to its popularity among users working with Python. To support Jupyter Notebooks, Microsoft’s Visual Studio Code (VS Code) was chosen because it has a lot more Jupyter Notebook users than JetBrains' PyCharm IDE. JetBrains had 15.9M users in 2023\footnote{\url{https://www.jetbrains.com/lp/annualreport-2023/}} and 11.4M concurrent users in 2024\footnote{\url{https://www.jetbrains.com/lp/annualreport-2024/}}, while VS Code has 82M users who have downloaded the Jupyter Notebook Extension for use on VS Code\footnote{\url{https://marketplace.visualstudio.com/items?itemName=ms-toolsai.jupyter}}.  Not to mention the 15.9M and 11.4M numbers for JetBrains comprise the total combined user of all JetBrains services.

As an illustrative example, Figure \ref{fig:example}, shows a step-by-step
scenario to automatically detect and fix the Data Leakage instance. Once the user installs \toolname from the VS Code marketplace and chooses the ``Data Leakage'' extension from the activity bar on the left, the user clicks on the gear icon on the top right to open the User Settings tab of the extension \ballnumber{1}. The user can then choose the preferred run mode from the dropdown menu, either ``Native''  \ballnumber{1.1} or ``Docker''  \ballnumber{1.2}, and returns to the main extension page by clicking the run icon. In the extension window, the user clicks ``Run Data Leakage Analysis'' to start the process \ballnumber{2}. Once the analysis is complete, the user receives a notification at the bottom right of VS Code \ballnumber{3}. After that, the user reviews the ``Leakage Overview'' tab in the bottom panel of VS Code, which displays a summary of detected leakages and provides a detailed table of instances \ballnumber{4}. Each instance can be examined by clicking on a row in the table. The user then navigates to a Data Leakage instance by selecting a row in the leakage instances table \ballnumber{5}. The selected leakage instance will be highlighted in the Jupyter Notebook file. The next step involves hovering over the highlighted line with the red error to reveal the ``Quick Fix'' option \ballnumber{6}. If the user selects ``Quick Fix'',  several potential solutions will be shown: (1) perform the manual ``Quick Fix'', (2) select the option ``Fix using Copilot'', or (3) select the option ``Ask Continue'  to perform AI-based Quick Fix (assuming that the user already installs the GitHub \textsc{Copilot} and \textsc{Continue} VS Code extensions to fix using \textsc{Copilot} and \textsc{Continue}, respectively) \ballnumber{7}. The Jupyter Notebook will be updated to remove the Data Leakage instance. In either case, the user will be asked to keep or cancel the proposed change \ballnumber{8}. 

\section{Study Design} 
\label{Section:methodology}

\subsection{Comparison between \toolname 1.0 and 2.0}
Our new VS Code extension introduces a Data Leakage tool specifically tailored for Jupyter Notebook files, unlike our previous PyCharm plugin, which was designed for Python files \cite{alomar2025leakagedetector}.  The previous version of the tool not only provides a graphical user interface to run and analyze leakages using the existing leakage analysis algorithm \cite{yang2022data} within Docker, but also provides quick fix solutions. Furthermore, when extracting data from the Data Leakage algorithm \cite{yang2022data}, it focused exclusively on parsing the generated CSV and Fact files to identify leakage information. However, after further analysis, it became evident that this method might lead to incorrect interpretations due to the raw and disjointed nature of these files. For instance, a CSV might list all the training and testing lines present in the user's file but does not indicate which pairs are actually related to each other. Understanding these relationships may then require cross-referencing with another generated file—yet which file exactly, and how to extract or align that information, is unclear and not clearly documented. Essentially, this task was akin to knowing the output of a very complex equation, but being given only the raw numbers and having to reverse-engineer the operations ourselves. Recognizing these limitations, we transitioned to directly parsing the generated HTML file. This file consolidates the final analysis output and clearly states not only which variables are involved in leakages, but also provides rich context such as the number of leakages detected, their exact locations, the relevant train and test sites, sources of leakage, and other usages of the same variable. This shift improved both the accuracy and reliability of our results, as it eliminated the need to manually piece together relationships from intermediate, ambiguous data sources. 

Additionally, our VS Code extension does the analysis on Jupyter Notebooks with a choice of running the core leakage analysis algorithm, natively or in Docker. Not only that, the extension also supports a better approach to perform the fix and has automatic CI/CD deployment, so we can easily work on extending the extension in the future. In summary, our VS Code extension had the following advantages:

\begin{itemize}
    \item \textbf{Conventional quick fix.} Our tool provides a less rudimentary solution to leakages inside the VS Code extension by providing fixes to some extra edge cases. In addition, it also provides an interface to show quick fix changes before and after a change called ``diff''. The user can choose to either discard or approve the changes to their Jupyter Notebook file caused by the quick fix solution in the main page of the extension. 

   \item \textbf{AI quick fix.}  VS Code has a wide variety of extensions including \textsc{Copilot} and \textsc{Continue}, which are two extensions that specialize in LLM integration with VS Code. Our VS Code extension is compatible with these extensions, and the user can utilize the existing AI quick fix within our extension once it is installed.

    \item \textbf{Docker \& native binaries.} The user can choose to use Docker for the Python leakage analysis program or use our distributed binaries (\ie binaries executable by the host operating system) version of the algorithm for all to analyze leakages. 

   \item \textbf{Table display for analyzing leakages.} Compared to its predecessor, our extension contains more information to display to the user such as ``Data Variable Name'', ``Model Variable Name'' and ``Method'' which could help to better understand the detected leakage instances.

\end{itemize}

The original leakage analysis algorithm is portable and flexible, but the user is required to install the dependency to run Docker on Python files. Thus, the need to convert Jupyter Notebooks to Python files is essential before conducting leakage analysis. Beyond that, it is challenging to get a glimpse of all of the leakages from the HTML file since the leakages are spread out in the HTML file.  This was what our PyCharm plugin went out to solve (\ie, provides a way for data scientists to have an overview of all of the leakages, and also a quick fix solution). The problem, however, is that it only works in PyCharm and, like the original leakage analysis algorithm, it only works with Python files. Because of that, the PyCharm plugin requires users to run Docker to use the plugin. To cope with this challenge, our VS Code extension solves the dependency issue by distributing binaries for each of the three major operating systems: Linux, Windows, and MacOS, and solves the problem of only working Python files by allowing Jupyter Notebooks to be used. Docker is no longer a necessity, but can still be used instead of the native binaries, and the user is provided with the option to choose their preferred method within the extension.  Our VS Code extension also provides a less naïve solution of quick fix compared to the PyCharm plugin. After running an analysis on the Jupyter Notebook file, the user may select the ``Quick Fix'' option for a given line of code to receive suggested edits that will remove the associated leakage instance. This allows users to quickly identify and correct instances of Data Leakage using the given corrections. In addition to the built-in quick fix methods, the extension also offers integration with GitHub \textsc{Copilot} or \textsc{Continue} to provide LLM-powered solutions to leakage instances. This creates a more dynamic system that provides users with multiple options to resolve the issues in their code. 
 Further, VS Code extension tables have more data than the PyCharm equivalent, so more analysis can be done in VS Code than in PyCharm. 

\subsection{Quick Fixes}
\label{sec:quick-fix}
Quick fixes are designed for convenience, helping developers maintain their workflow without having to manually write every minor correction. However, they are typically rudimentary solutions, they address surface-level symptoms rather than deeply understanding the context of the code. As a result, while quick fixes can be very helpful for simple issues (such as missing imports, basic syntax errors, or common refactoring tasks), they are not always guaranteed to work and may require manual verification or adjustment afterward. Next, we introduce the Data Leakage types with quick fixes using the code example provided by Yang \etal \cite{yang2022data}.

\subsubsection{Overlap Leakage} Overlap leakages occur when some or all test data is mistakenly used during training or hyper-parameter tuning. A typical example is where an overlap leakage arises, when the variable \texttt{X}, which contains both training and test data, is used to train the ridge model. A proper fix would involve replacing \texttt{X} with \texttt{X\_train}, the variable returned by the \texttt{train\_test\_split} method that contains only the training data. Our quick fix algorithm automates this repair. The following steps are taken: 
\begin{enumerate}
    \item Find the latest call to \texttt{train\_test\_split} in the code. 
    \item Identify the \texttt{X\_train} and \texttt{y\_train} outputs (typically the first and third elements of the returned tuple from scikit-learn’s \texttt{train\_test\_split}). 
    \item Replace the parameters of the offending fitting method with the correct \texttt{X\_train} and \texttt{y\_train} variables. 
\end{enumerate}
After applying the quick fix, the code would look like Listing \ref{Listing:overlap}: 

\begin{lstlisting}[caption=A quick fix applied to Overlap Leakage., label=Listing:overlap, numbers=none, firstnumber = last, escapeinside={(*@}{@*)}]
import pandas as pd
from sklearn.feature_selection import (SelectPercentile, chi2)
from sklearn.model_selection import (LinearRegression, Ridge)

X_0, y = load_data()

select = SelectPercentile(chi2, percentile=50)
select.fit(X_0)
X = select.transform(X_0)

X_train, X_test, y_train, y_test = train_test_split(X, y)
lr = LinearRegression()
lr.fit(X_train, y_train)
lr_score = lr.score(X_test, y_test)

ridge = Ridge()
ridge.fit(X_train, y_train)
ridge_score = ridge.score(X_test, y_test)

final_model = lr if lr_score > ridge_score else ridge

\end{lstlisting}


\subsubsection{Preprocessing Leakage} 
Preprocessing leakages happen when training and test data are preprocessed together, allowing information from the test set to improperly influence the training set. A typical example a preprocessing leakage occurs when the variable \texttt{X\_0}, which contains both training and test data, is transformed. This causes the resulting variable \texttt{X}—which is later split—to have its \texttt{X\_train} inadvertently influenced by \texttt{X\_test}. A proper fix is to move the preprocessing and feature selection steps after the \texttt{train\_test\_split}. This ensures that only training data is used to learn transformations, preserving the integrity of the evaluation. Our quick fix algorithm handles this correction by following these steps: 
\begin{enumerate}
    \item Locate the latest call to the \texttt{train\_test\_split} method. 
     \item Identify the \texttt{X\_train} and \texttt{X\_test} outputs of the split. 
\item Find the lines where preprocessing occurs. 
\item Move the \texttt{train\_test\_split} call upward, placing it before the preprocessing. 
\item Assign the outputs of \texttt{train\_test\_split} to temporary variables (\eg \texttt{X\_train\_0} and \texttt{X\_test\_0}). 
\item Apply all preprocessing and transformation methods to these temporary variables. 
\item Have the results of preprocessing be the original \texttt{X\_train} and \texttt{X\_test} variables. 
\end{enumerate}
In essence, the quick fix introduces intermediate variables that ``absorb'' the transformations, and their processed versions are then assigned back to the expected names.  After applying the quick fix, the code would look like Listing \ref{Listing:preprocessing}:

\begin{lstlisting}[caption=A quick fix applied to Preprocessing Leakage., label=Listing:preprocessing, numbers=none, firstnumber = last, escapeinside={(*@}{@*)}]
import pandas as pd
from sklearn.feature_selection import (SelectPercentile, chi2)
from sklearn.model_selection import (LinearRegression, Ridge)

X_0, y = load_data()

X_train_0, y_train, X_test_0, y_test = train_test_split(X, y)
select = SelectPercentile(chi2, percentile=50)
select.fit(X_train_0)
X_train = select.transform(X_train_0)
X_test = select.transform(X_test_0)

lr = LinearRegression()
lr.fit(X_train, y_train)
lr_score = lr.score(X_test, y_test)

ridge = Ridge()
ridge.fit(X, y)
ridge_score = ridge.score(X_test, y_test)

final_model = lr if lr_score > ridge_score else ridge
       

\end{lstlisting}


\subsubsection{Multi-test Leakage}
Multi-test leakages occur when the same test data is repeatedly used for evaluation, blurring the line between validation and true testing. A common example where the variable \texttt{X\_test} is evaluated multiple times across both the \texttt{lr} and ridge models, leading to artificial inflation of performance metrics. A standard solution is to introduce a new independent test set that has not been reused or influenced by previous evaluations. Our quick fix algorithm addresses multi-test leakages using the following steps: 
\begin{enumerate}
    \item Identify the reused test variable (\eg \texttt{X\_test}) that is evaluated multiple times across different models. 
   \item Insert a placeholder loading method, such as \texttt{load\_test\_data()}, to simulate retrieving a new, independent test dataset. 
\item Create new test variables derived from the original test variable (\eg \texttt{X\_X\_test\_new\_0} and \texttt{y\_X\_test\_new}) through the \texttt{load\_test\_data()} function. 
\item Select a user-initialized model that performs transformations (\eg a scaler or preprocessor). 
\item Apply the model’s transformation method (specifically looking for the transform keyword) to \texttt{X\_X\_test\_new\_0}, resulting in a transformed variable \texttt{X\_X\_test\_new}. 
\item Select the last user-initialized model capable of evaluation (\eg a classifier or regressor). 
\item Evaluate the selected model using the newly created \texttt{X\_X\_test\_new} and \texttt{y\_X\_test\_new} variables, ensuring that the evaluation does not reuse the original, potentially tainted test set. 
\end{enumerate}

After applying the quick fix, the code would look like Listing \ref{Listing:multitest}: 

\begin{lstlisting}[caption=A quick fix applied to Multi-test Leakage., label=Listing:multitest, numbers=none, firstnumber = last, escapeinside={(*@}{@*)}]
import pandas as pd
from sklearn.feature_selection import (SelectPercentile, chi2)
from sklearn.model_selection import (LinearRegression, Ridge)

X_0, y = load_data()

select = SelectPercentile(chi2, percentile=50)
select.fit(X_0)
X = select.transform(X_0)

X_train, X_test, y_train, y_test = train_test_split(X, y)

lr = LinearRegression()
lr.fit(X_train, y_train)
lr_score = lr.score(X_test, y_test)

ridge = Ridge()
ridge.fit(X, y)
ridge_score = ridge.score(X_test, y_test)

final_model = lr if lr_score > ridge_score else ridge
X_X_test_new_0, y_X_test_new = load_test_data()
X_X_test_new = select.transform(X_X_test_new_0)
final_model.score(X_X_test_new_0, y_X_test_new)

\end{lstlisting}


\subsection{\toolname without Docker}
The Docker-free solution (\ie local installation) is a program that runs the leakage algorithm in native mode. The main advantage of this is that running the leakage algorithm will consume significantly less memory and CPU usage over a running Docker container. The solution is packaged as a single zip file that contains all the required dependencies. This allows the installation process of the program to be as simple as possible for the user. These are the main components of the Docker-free solution: 
\begin{enumerate}
    \item Pyright Module: An npm package that runs with Node and is responsible for outputting important type information of the Jupyter Notebook code. 
    \item Python Code: This is the main code that is responsible for collecting and parsing program input as well as calling all the other components as sub-processes. 
    \item Main.dl Executable: This is the executable that runs the leakage algorithm from the input supplied by the program and outputs a set of csv files. 
\end{enumerate}

\subsection{\toolname with Docker}
The Docker solution runs the leakage algorithm within a Docker container environment. The main advantage of this approach is that it provides a consistent and isolated environment that works with different operating systems and configurations. The solution only requires the user to download and run Docker desktop. All the complexities of downloading images and running containers are handled by the application, making the process straightforward for users. These are the main components of the Docker solution: 
\begin{enumerate}
    \item Docker Image: A pre-configured environment containing all required dependencies, libraries, and executables needed to run the leakage algorithm. 
    \item Container Runtime: Managed by the application, this handles the creation and execution of the Docker container that runs the leakage algorithm. 
    \item Volume Mounting: Facilitates access to local files from within the container, allowing seamless input and output between the host system and the containerized environment. 
\end{enumerate}

\subsection{Table Panel} 
After running Data Leakage analysis successfully, the two tables (Leakage Summary and Leakage Instances) in VS Code’s panel will be populated with values. Leakage Summary has 3 rows to represent Overlap, Preprocessing, and Multi-test with their respective frequency/count on the total error in a specific Notebook file. Like the Leakage Summary table, once the user runs the leakage tool, the Leakage Instances table will be populated with the following information: Type, General Cause, Cell, Line, Model Variable Name, Data Variable Name, and Method, which gives the user more context about the detected leakages. 

\section{Preliminary Evaluation}
\label{Section:Result}

We evaluated the effectiveness of \toolname in detecting Data Leakage and fixing it. Because \toolname is developed using the leakage static analysis tool as a foundation, we did not assess the detection accuracy of the tool separately. Instead, we considered the performance reported by Yang \etal \cite{yang2022data} to be valid for our purposes as well.

We utilize the existing publicly available dataset \cite{yang2022data} that contains Data Leakage of different types to evaluate our tool. We started by selecting 31 Python files, converting them to Jupyter Notebooks, and manually analyzing the existing annotated sets and running them using our VS Code extension. The results of the analysis show the leakage detected by \toolname, and we use our quick fix, as well as the LLM-driven approach, \ie \textsc{Copilot} and \textsc{Continue}. VS Code extensions.

In addition to our internal evaluation, we invited 3 participants representing undergraduate and graduate students from Stevens
Institute of Technology to use any of the files containing files when trying out our extension. All participants volunteered for the experiment. Overall, the participants were very satisfied with the setup, documentation, execution time, and quick fixes of our VS Code extension tool.

\section{Related Work}
\label{Section:RelatedWork}

 Yang \etal \cite{yang2022data} proposed a static analysis algorithm to detect common forms of Data Leakage in data science code. Bouke and Abdullah \cite{bouke2023empirical} examined how Data Leakage during data preprocessing affects the dependability of machine learning-based intrusion detection systems. Lopez \etal \cite{lopez2024inter} explored inter-dataset code duplication and its impact on evaluating LLMs across diverse software engineering tasks. Their results show that open-source models such as \textsc{CODEBERT}, and \textsc{GRAPHCODEBERT} could be affected by inter-dataset duplication.
 Kery \etal \cite{kery2018story}  
 interviewed 21 data scientists to study coding behaviors. Their findings stimulate the development of novel designs for interacting with notebook cells, facilitating tasks such as browsing history, debugging, and more, which could enhance the efficiency of literate programming in supporting data science endeavors.  Drobnjakovi{\'c} \etal \cite{drobnjakovic2024abstract} developed a static analysis based on abstract interpretation to verify the absence of Data Leakage. Their approach was incorporated into the NBLyzer framework and was evaluated for its effectiveness and accuracy using 2111 Jupyter Notebooks sourced from the Kaggle competition platform. Venkatesh \etal \cite{venkatesh2023enhancing} introduced HeaderGen, a tool that autonomously labels code cells with categorical markdown headers using a taxonomy for machine learning operations and classifies and displays function calls based on this classification. 

\section{Conclusion and Future Work}
\label{Section:Conclusion}

We develop \toolname, a VS Code extension that supports the detection and correction of Data Leakage. It offers several features that include a docker-free solution, less naive quick fixes, and an LLM-driven
approach that guides ML developers toward best practices for
building ML pipelines. It is designed to support Jupyter Notebook files. 

\section{Acknowledgments}
\label{sec:ack}
We would like to thank the authors of the leakage static analysis tool \cite{yang2022data} for publicly providing it.


\bibliographystyle{abbrv}

\bibliography{IEEEabrv,sample-base}

\begin{thebibliography}{1}

\bibitem{alomar2025leakagedetector}
E.~A. AlOmar, C.~DeMario, R.~Shagawat, and B.~Kreiser.
\newblock Leakagedetector: An open source data leakage analysis tool in machine learning pipelines.
\newblock pages 844--849, 2025.

\bibitem{bouke2023empirical}
M.~A. Bouke and A.~Abdullah.
\newblock An empirical study of pattern leakage impact during data preprocessing on machine learning-based intrusion detection models reliability.
\newblock {\em Expert Systems with Applications}, 230:120715, 2023.

\bibitem{drobnjakovic2024abstract}
F.~Drobnjakovi{\'c}, P.~Suboti{\'c}, and C.~Urban.
\newblock An abstract interpretation-based data leakage static analysis.
\newblock In {\em International Symposium on Theoretical Aspects of Software Engineering}, pages 109--126. Springer, 2024.

\bibitem{kery2018story}
M.~B. Kery, M.~Radensky, M.~Arya, B.~E. John, and B.~A. Myers.
\newblock The story in the notebook: Exploratory data science using a literate programming tool.
\newblock In {\em Proceedings of the 2018 CHI conference on human factors in computing systems}, pages 1--11, 2018.

\bibitem{lopez2024inter}
J.~A.~H. L{\'o}pez, B.~Chen, M.~Saad, T.~Sharma, and D.~Varr{\'o}.
\newblock On inter-dataset code duplication and data leakage in large language models.
\newblock {\em IEEE Transactions on Software Engineering}, 2024.

\bibitem{venkatesh2023enhancing}
A.~P.~S. Venkatesh, J.~Wang, L.~Li, and E.~Bodden.
\newblock Enhancing comprehension and navigation in jupyter notebooks with static analysis.
\newblock In {\em 2023 IEEE international Conference on software analysis, evolution and reengineering (SANER)}, pages 391--401. IEEE, 2023.

\bibitem{yang2022data}
C.~Yang, R.~A. Brower-Sinning, G.~Lewis, and C.~K{\"a}stner.
\newblock Data leakage in notebooks: Static detection and better processes.
\newblock In {\em Proceedings of the 37th IEEE/ACM International Conference on Automated Software Engineering}, pages 1--12, 2022.

\end{thebibliography}

\end{document}